\documentclass{elsart}
\usepackage{epsfig,amstex}
\renewcommand{\vec}{\boldsymbol}

\newcommand{\cal}{\mathcal}

\newcommand{\lsim}{
 \mathrel{\setbox0=\hbox{$<$}\raise0.6ex\copy0\kern-\wd0
 \lower0.65ex\hbox{$\sim$}}}

\newcommand{\gsim}{
 \mathrel{\setbox0=\hbox{$>$}\raise0.6ex\copy0\kern-\wd0
 \lower0.65ex\hbox{$\sim$}}}
 
\newcommand{\T}[1]{{\mathrm{#1}}}

\begin{document}

\thispagestyle{empty}
\vspace*{-2cm}
\begin{flushright}
\bf 
TUM/T39-98-28
\end{flushright}

\bigskip 
\bigskip 
\bigskip

\begin{center}
{\Large\bf Note on Shadowing and Diffraction in \\[0.3cm] 
Deep-Inelastic Lepton Scattering $^{*)}$}

\vspace{2.cm}

{\large G.Piller$^{a}$, L.~Ferreira$^{b}$, W. Weise$^{a}$} 


\vspace{2.cm}

$^{(a)}$ Physik Department, Technische Universit\"{a}t M\"{u}nchen,
D-85747 Garching, Germany \\[0.2cm]
$^{(b)}$ Departamento de Fisica and CFIF-Edificio Ciencia, 
        Instituto Superior Tecnico \\ Avenida Rovisco Pais, 
        1096 Lisboa, Portugal

\vspace*{3cm}

{\bf Abstract}

\bigskip
\begin{minipage}{15cm}
We discuss the close relation between  shadowing in deep-inelastic 
lepton-nucleus scattering and diffractive photo- and leptoproduction 
of hadrons from free nucleons. 
We show that the magnitude of nuclear shadowing at small 
Bjorken-$x$, as measured by the E665 and NMC collaborations, 
is directly related to HERA data on the amount of 
diffraction in the scattering  from free nucleons.

\end{minipage}

\end{center} 



\vspace*{3.cm}

------------------------------------
\\[0.1cm]
\noindent $^{*}$) Work supported in part by BMBF  and DAAD.

\newpage

In recent years diffractive  photo- and leptoproduction processes 
have received  much attention. 
This interest has been initiated mainly by the large  amount of 
new data from the HERA electron-proton collider 
(for reviews see \cite{Cartiglia:1996xv,Gallo:1997ez}).   
While diffraction from free nucleons is an interesting subject 
all by itself, it also plays a major role in the shadowing phenomena 
observed in deep-inelastic lepton scattering from nuclei. 
Here plenty of data have become available in the last decade from 
high precision experiments at 
CERN (NMC) \cite{Amaudruz:1995tq,Arneodo:1995cs}
and FNAL (E665) 
\cite{Adams:1992nf,Adams:1995is}.

In this note we exploit the close relation between shadowing 
and diffraction \cite{Gribov:1970}. 
The connection of these two phenomena 
is well known and used frequently in the literature. 
Nevertheless, it is instructive to see  that 
the magnitude  of nuclear  shadowing in  deep-inelastic lepton 
scattering is determined in a simple and basic way 
by the relative amount of diffraction observed  
in the high-energy scattering from free nucleons.

In deep-inelastic lepton scattering  from nuclei 
shadowing  occurs  at 
small values of the Bjorken scaling variable $x=Q^2/2 p\cdot q$, 
where $p^{\mu}$ and $q^{\mu}$ are the nucleon and photon 
four-momenta, respectively, and $Q^2 = -q^2$. 
In the following we choose the photon momentum in the $\hat z$-direction, 
i.e. $\vec q = (\vec 0_{\perp},q_z)$.  
At $x<0.1$ hadronic components of the photon field can 
interact coherently with several nucleons in the nuclear target. 
Destructive interference of  multiple scattering 
amplitudes leads to a depletion (shadowing) 
of  nuclear structure functions $F_2^{\T A}$ 
as compared to $\T A$ times the one for free nucleons, $F_2^{\T N}$. 
The same  observation holds, of course,  for the 
virtual photon-nucleus cross section 
$ \sigma_{\gamma^* \T A} = (4 \pi^2 \alpha/Q^2) \, F_2^A$.

This cross section can be separated 
into a piece which accounts for the incoherent scattering from 
individual nucleons, and a correction due to the coherent interaction 
with several nucleons, 
$\sigma_{\gamma^* \T A} = {\T A} \, \sigma_{\gamma^* \T N} + 
\delta  \sigma_{\gamma^* \T A}$. 
The multiple scattering correction  $\delta  \sigma_{\gamma^* \T A}$ 
is dominated by contributions which involve two nucleons. 
In this note we restrict ourselves to this so-called 
double scattering correction\footnote{ 
Corrections due to triple scattering amount typically to less than 
$30\%$ of the  double scattering term \cite{Piller:1995kh}.}.
It is controlled by the diffractive production of hadrons 
from single nucleons and their subsequent rescattering. 
This process, illustrated in Fig.1, is described by 
the well known relation \cite{Gribov:1970}:
\begin{eqnarray} \label{eq:ds_A}
\delta  \sigma_{\gamma^* \T A} &\approx& 
- 8\pi \int_{4 m_{\pi}^2}^{W^2} dM_{\T X}^2 
\int d{^2 b} \int_{-\infty}^{+\infty} dz_1 
\int_{z_1}^{+\infty} dz_2 \,
\rho_{\T A}(\vec b, z_1) \rho_A(\vec b, z_2) \,
\nonumber \\  
&&\hspace*{3cm}\times
\cos\left[ (z_2 - z_1)/\lambda \right] 
\left. \frac{d^2\sigma^{diff}_{\gamma^{*} {\T N}}}{dM_{\T X}^2 dt}
\right|_{t\approx 0}.  
\end{eqnarray}
A state $\T X$ with invariant mass $M_{\T X}$ is 
produced diffractively in the 
process  $\gamma^* \T N \rightarrow \T{X N}$, 
with the nucleon located at $(\vec b, z_1)$. 
The upper limit of $M_{\T X}$ is determined by the available 
$\gamma^* {\T N}$ center-of-mass energy $W$. 
The hadronic excitation propagates at fixed impact parameter $\vec b$
and then interacts  with  a second nucleon at $z_2$. 
The underlying basic mechanism, i.e. diffraction from a single nucleon, 
is described  by the cross section 
$d^2\sigma^{diff}_{\gamma^{*} {\T N}}/{dM_{\T X}^2 dt}$ 
taken in the forward direction, $t = (p-p')^2 \approx 0$, 
where $p'$ is the four-momentum of the scattered nucleon.
The product of nuclear densities,  
$\rho_{\T A} (\vec b,z_1)\,\rho_{\T A} (\vec b,z_2)$, 
accounts for the probability to find two nucleons in the target 
at the same impact parameter. 
We use the normalization $\int d^3 r\,\rho_{\T A} (\vec r) = A$.
Finally, the  $\cos [ (z_2 - z_1)/\lambda]$ 
factor in (\ref{eq:ds_A}) implies that only diffractively 
excited hadrons with a longitudinal propagation length larger than 
the average nucleon-nucleon distance in the target, 
\begin{equation}
\lambda = \frac{2 \nu}{Q^2 + M_{\T X}^2} \, > d \approx 2\,\rm{fm},
\end{equation}
can contribute significantly to double scattering.

Equation (\ref{eq:ds_A}) has been applied in several investigations 
of nuclear shadowing, using different models for the diffractive 
photoproduction cross section 
(see e.g. 
\cite{Piller:1995kh,Kwiecinski:1988ys,Frankfurt:1989zg,Nikolaev:1991ja,%
Melnitchouk:1993vc,Piller:1997ny,Barone:1997ij,Capella:1997mn}). 
A simple estimate of nuclear shadowing 
at small Bjorken-$x$ can, however, already be obtained 
by just looking at the relative amount 
of diffractive production in deep-inelastic scattering  from free nucleons 
which is  known from recent experiments at HERA. 


\begin{figure}[t]
\begin{center} 
\epsfig{file=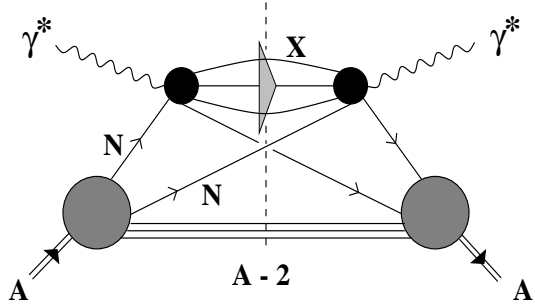,height=40mm,width=70mm}
\end{center}
\caption[...]{Double scattering contribution 
to deep-inelastic charged lepton scattering from nuclei. 
\vspace*{0.5cm}
}
\label{fig:mult_double} 
\end{figure}


For $x\ll 0.1$, the coherence lengths $\lambda$ of the hadronic states 
which dominate diffractive production in eq.(\ref{eq:ds_A}), 
exceed the diameter of the target nucleus.  
In the limit $\lambda \rightarrow \infty$ we find: 
\begin{equation} \label{eq:appr}
\delta  \sigma_{\gamma^* \T A} \simeq - 8\pi \,B\,
\sigma^{diff}_{\gamma^* {\T N}}
\int d{^2 b} \int_{-\infty}^{+\infty} dz_1 
\int_{z_1}^{+\infty} dz_2 \,
\rho_{\T A}(\vec b,z_1) \,\rho_{\T A}(\vec b, z_2).  
\end{equation}
The slope parameter $B$ describes the $t$-dependence of the 
diffractive production cross section
\begin{equation} \label{eq:def_diff_cross_approx} 
\frac{d^2\sigma^{diff}_{\gamma^{*} {\T N}}}{dM_{\T X}^2 dt} 
=
e^{-{B} |t|}\,\left.
\frac{d^2\sigma^{diff}_{\gamma^{*} {\T N}}}{dM_{\T X}^2 dt}
\right|_{t = 0}\,.
\end{equation}
Recent HERA data on the integrated diffractive leptoproduction 
cross section are well described using  
$B \simeq 7$ GeV$^{-2}$ \cite{Breitweg:1998aa}. 
In the diffractive production of low mass vector mesons  
($\rho, \omega$ and $\phi$) from nucleons, a range of values 
$B \simeq (4 \dots 10)$ GeV$^{-2}$ has been found, depending on 
$Q^2$ and on the  incident photon energy 
(for reviews see e.g. \cite{Crittenden:1997yz}).

For the nuclear densities in eq.(\ref{eq:appr}) we use Gaussian, 
\begin{equation} \label{eq:Gauss}
\rho_{\T A}(\vec{r})
= A \,\left(\frac{3}{2 \,\pi\,\langle r^2 \rangle_{\T A} }\right)^{3/2}  \,
\exp \left( - 
\frac{3\,\vec{r}\,^2}{2\,\langle r^2 \rangle_{\T A}} 
\right),
\end{equation}
and square-well parametrizations,  
\begin{equation}
\rho_{\T A}(\vec{r}) = 
\left\{ \begin{array}{l}
A \frac{3}{4\pi} 
\left(\frac{3}{5 \,\langle r^2 \rangle_{\T A} }\right)^{3/2}  
\quad 
\mbox{for}
\quad r < \sqrt{\frac{5}{3}} \,\langle r^2 \rangle_{\T A}^{1/2}, 
\nonumber \\
0  \hspace*{3cm}\mbox{otherwise},
\end{array}\right.
\end{equation}
with  the mean square radius $\langle r^2 \rangle _{\T A} =  
\int d^3r \,r^2 \,\rho_{\T A}(r)/A$. 
For both cases the shadowing ratio 
$R_{\T A} = 
\sigma_{\gamma^* {\T A}}/
{A \sigma_{\gamma^* {\T N}}}$
is easily worked out:
\begin{equation} \label{eq:shad_est}
R_{\T A} \simeq 1 -  {\cal C} \,A\,
\left(\frac{B}{\langle r^2 \rangle _{\T A}} \right)
\frac{\sigma^{diff}_{\gamma^* {\T N}}}{\sigma_{\gamma^* {\T N}}}.
\end{equation}
For Gaussian nuclear densities one finds ${\cal C}=3$, while 
${\cal C}=2.7$ in the  square-well case.

To estimate the magnitude of nuclear shadowing as 
observed by  NMC and E665 
one needs to know the diffraction part of the free nucleon 
cross section for comparable  kinematic conditions. 
Recent data on diffractive production   are available from  HERA
\cite{Derrick:1994dt,Aid:1995bz,Adloff:1997mi}
for incident real photons.  
It has been found that 
diffractive processes, which leave the target proton intact,  amount 
to approximately $20\%$ of the total photon-nucleon cross section 
at  center-of-mass energies $W \sim 200$ GeV.
About half of the diffractive events are due to 
vector meson production. 
In leptoproduction at large $Q^2\sim 10$ GeV$^2$,
on the other hand, diffractive production  reduces 
to about  $10\%$ of the total deep-inelastic scattering cross section 
\cite{Derrick:1996ma}.

In fixed target lepton scattering experiments the average squared 
momentum transfer $\overline Q^2$ decreases for decreasing $x$. 
For the NMC and E665 experiments  one finds, for example,  
$\overline Q^2 \lsim 0.4$  GeV$^2$ at $x\lsim 10^{-3}$ 
\cite{Arneodo:1995cs,Adams:1995is}.   
In this high-energy region it is legitimate to assume 
Regge behavior for the energy dependence of diffraction 
(see e.g. \cite{Adloff:1997mi}). 
At $x\sim 10^{-4}$ the center-of-mass energies 
in the NMC and E665 measurements are 
$W \sim 15$ GeV \cite{Arneodo:1995cs} and 
$W \sim 25$ GeV  \cite{Adams:1995is}, respectively. 
For these energies the relative amount 
of diffraction  reaches about  $60\%$ of 
the value found at HERA. 
Note that this agrees  within $25\%$ with 
an  analysis of earlier data on diffraction from 
FNAL \cite{Chapin:1985}. 

In our estimate (\ref{eq:shad_est}) we use $B\approx 8$ GeV$^{-2}$ 
for the slope parameter. 
This is a  reasonable average of the corresponding 
values extracted from diffractive photoproduction of low mass vector 
mesons \cite{Crittenden:1997yz,Chapin:1985} and high mass states 
\cite{Breitweg:1998aa,Chapin:1985}.
Using furthermore 
${\sigma^{diff}_{\gamma^* {\T N}}}/{\sigma_{\gamma^* {\T N}}}\simeq 0.1$     
the magnitude of $R_{\T A}$ comes out in very good agreement with
experimental values as shown in table \ref{tab}.
\begin{table}[h]
\begin{center}
\begin{tabular}{| c | c | c | c | c | }
\hline 
                  & $^6$Li  & $^{12}$C & $^{40}$Ca & $^{131}$Xe \\ \hline     
 $R_{\T A}$       & $0.93$  & $0.84$   & $0.73$    & $0.65$    \\ \hline
 $R_{\T A}^{exp.}$& $0.94 \pm 0.07$  & $0.87 \pm 0.10$   & $0.77 \pm 0.07$ 
                   &  $0.67 \pm 0.09$ \\ \hline
\end{tabular}
\caption{
The shadowing ratio $R_{\T A}$ estimated according to eq.(\ref{eq:shad_est}) 
in comparison with experimental data for various nuclei. 
Here square-well nuclear densities have been used.
The data are taken from 
ref.\cite{Adams:1992nf,Adams:1995is,Arneodo:1995cs} at the smallest 
kinematically accessible values of the Bjorken variable $x$ 
($x \simeq 10^{-4}$). 
}
\label{tab}
\end{center}
\end{table}
This estimate confirms in a simple and basic way that shadowing 
in nuclear deep-inelastic scattering 
is governed by the coherent interaction of diffractively 
produced states with several nucleons in the target nucleus.

Note that shadowing and diffraction are also linked 
in  deep-inelastic charged current interactions. 
Nuclear shadowing and diffraction have been observed 
here too, although with large experimental errors 
(for a review see e.g.\cite{Kopeliovich:1993ym}). 
A detailed theoretical investigation 
of shadowing effects in neutrino-nucleus scattering 
has been carried out recently in ref.\cite{Boros:1998mt}. 
It is found that 
shadowing in charged lepton and neutrino scattering is 
of  similar magnitude.  
Then 
the relative amount of diffraction also is expected to 
be roughly the same.

To summarize: based on the simple geometric relation 
(\ref{eq:shad_est}) 
we have illustrated how the magnitude 
of nuclear shadowing at $x\simeq 10^{-4}$, as measured by the 
NMC and E665 collaborations, 
is directly and very easily 
connected to the relative amount of diffraction 
from free nucleons which has been determined recently at HERA.
This close connection emphasizes the role of diffraction in 
nuclear deep-inelastic scattering at small Bjorken-$x$, 
and eq.(\ref{eq:shad_est}) proves useful for practical estimates 
of the leading coherent double scattering effect.


\begin{thebibliography}{10}

\bibitem{Cartiglia:1996xv}
N. Cartiglia,
\newblock (1996), hep-ph/9703245.

\bibitem{Gallo:1997ez}
E. Gallo,
\newblock (1997), hep-ex/9710013.

\bibitem{Amaudruz:1995tq}
NMC, P. Amaudruz et~al.,
\newblock Nucl. Phys. B441 (1995) 3.

\bibitem{Arneodo:1995cs}
NMC, M. Arneodo et~al.,
\newblock Nucl. Phys. B441 (1995) 12.

\bibitem{Adams:1992nf}
E665, M.R. Adams et~al.,
\newblock Phys. Rev. Lett. 68 (1992) 3266.

\bibitem{Adams:1995is}
E665, M.R. Adams et~al.,
\newblock Z. Phys. C67 (1995) 403.

\bibitem{Gribov:1970}
V.N. Gribov,
\newblock Sov. Phys. JETP 30 (1970) 709.

\bibitem{Piller:1995kh}
G. Piller, W. Ratzka and W. Weise,
\newblock Z. Phys. A352 (1995) 427.

\bibitem{Kwiecinski:1988ys}
J. Kwiecinski and B. Badelek,
\newblock Phys. Lett. 208B (1988) 508.

\bibitem{Frankfurt:1989zg}
L.L. Frankfurt and M.I. Strikman,
\newblock Nucl. Phys. B316 (1989) 340.

\bibitem{Nikolaev:1991ja}
N.N. Nikolaev and B.G. Zakharov,
\newblock Z. Phys. C49 (1991) 607.

\bibitem{Melnitchouk:1993vc}
W. Melnitchouk and A.W. Thomas,
\newblock Phys. Lett. B317 (1993) 437.

\bibitem{Piller:1997ny}
G. Piller, G. Niesler and W. Weise,
\newblock Z. Phys. A358 (1997) 407.

\bibitem{Barone:1997ij}
V. Barone and M. Genovese,
\newblock Phys. Lett. B412 (1997) 143.

\bibitem{Capella:1997mn}
A. Capella et~al.,
\newblock (1997), hep-ph/9702241.

\bibitem{Breitweg:1998aa}
ZEUS, J. Breitweg et~al.,
\newblock Eur. Phys. J. C1 (1998) 81.

\bibitem{Crittenden:1997yz}
J.A. Crittenden,
\newblock Tracts in Modern Physics, Volume 140, Springer 1997.

\bibitem{Derrick:1994dt}
ZEUS, M. Derrick et~al.,
\newblock Z. Phys. C63 (1994) 391.

\bibitem{Aid:1995bz}
H1, S. Aid et~al.,
\newblock Z. Phys. C69 (1995) 27.

\bibitem{Adloff:1997mi}
H1, C. Adloff et~al.,
\newblock Z. Phys. C74 (1997) 221.

\bibitem{Derrick:1996ma}
ZEUS, M. Derrick et~al.,
\newblock Z. Phys. C70 (1996) 391.

\bibitem{Chapin:1985}
T.J. Chapin et~al.,
\newblock Phys. Rev. D31 (1985) 17.

\bibitem{Kopeliovich:1993ym}
B.Z. Kopeliovich and P. Marage,
\newblock Int. J. Mod. Phys. A8 (1993) 1513.

\bibitem{Boros:1998mt}
C. Boros, J.T. Londergan and A.W. Thomas,
\newblock (1998), hep-ph/9804410.

\end{thebibliography}
\end{document}